\documentclass[aps,prl,twocolumn,amsmath,amssymb,superscriptaddress]{revtex4-1}

\usepackage{graphics}
\usepackage{dcolumn}% Align table columns on decimal point
\usepackage{bm}% bold math
\usepackage[caption=false]{subfig} 
\usepackage{color}
\usepackage{tikz}
\usepackage{subfig}
\usepackage[titles,subfigure]{tocloft}
\usepackage{soul}

\usepackage[titletoc,toc,title]{appendix}

\begin{document}

%\title{\color{blue}Spontaneous spin separation in a fractional topological insulator with bi-holomorphic Landau levels}
%\title{\color{blue}Spontaneous symmetry breaking in a fractional topological insulator with spin-dependent holomorphicity}
%\title{\color{blue}Fractional topological insulator with spin-dependent holomorphicity}
%\title{\color{blue}Spin-holomorphic fractional topological insulator}
\title{Spin separation due to an inherent spontaneous symmetry breaking of the fractional topological insulator}

\author{Sutirtha Mukherjee}
\affiliation{Quantum Universe Center, Korea Institute for Advanced Study, Seoul 02455, Korea}
\author{Kwon Park}
\email[Electronic address:$~~$]{kpark@kias.re.kr}
\affiliation{Quantum Universe Center, Korea Institute for Advanced Study, Seoul 02455, Korea}
\affiliation{School of Physics, Korea Institute for Advanced Study, Seoul 02455, Korea}

\date{\today}

\begin{abstract}
Motivated by the close analogy with the fractional quantum Hall states (FQHSs), fractional Chern insulators (FCIs) are envisioned as strongly correlated, incompressible states emerging in a fractionally filled, (nearly) flat band with non-trivial Chern number.  
Built upon this vision, fractional topological insulators (FTIs) have been proposed as being composed of two independent copies of the FCI with opposite Chern numbers for different spins, preserving the time-reversal symmetry as a whole. 
An important question is if the correlation between electrons with different spins can be really ignored.   
To address this question, we investigate the effects of correlation in the presence of spin-dependent holomorphicity, i.e., electrons of one spin species reside in the holomorphic lowest Landau level, while those of the other in the antiholomorphic counterpart.  
By constructing and performing exact diagonalization of an appropriate model Hamiltonian, here, we show that generic, strongly correlated, fractionally filled states with spin-dependent holomorphicity cannot be described as two independent copies of the FQHS, suggesting that FTIs in the lattice cannot be described as those of the FCI either.
Fractionally filled states in this system are generally compressible except at half filling, where an insulating state called the half-filled spin-holomorphic FTI occurs. 
It is predicted that the half-filled spin-holomorphic FTI is susceptible to an inherent spontaneous symmetry breaking, leading to the spatial separation of spins. 
\end{abstract}

\maketitle

Envisaged to occur in strongly correlated systems in a (nearly) flat band with non-trivial Chern number~\cite{Tang11,Sun11}, fractional Chern insulators (FCIs)~\cite{Neupert11_PRL,Sheng11,Wang11,Regnault11,Qi11,Wu12} are the lattice analogs of the fractional quantum Hall states (FQHSs)~\cite{Tsui82}. 
The analogy between FQHSs and FCIs can be made concrete by constructing the basis function mapping between the lowest-Landau-level wave functions and the so-called hybrid Wannier functions, which are localized in one direction, but extended in another~\cite{Qi11,Wu12}.  
Fractional topological insulators (FTIs)~\cite{Bernevig06_PRL,Levin09,Maciejko10,Santos11,Levin11,Lu12,Chen12,Levin12,Repellin14,Stern16} are distinguished from FQHSs and FCIs in the sense that the former preserve the time-reversal symmetry, while the latter do not.  
Conceptually, a FTI can be constructed by combining two independent copies of the FQHS or FCI with opposite Chern numbers for different spins~\cite{Levin09,Santos11,Lu12,Chen12,Repellin14}. 
Let us call this state the independent bipartite FTI.

An important question is if the correlation between electrons with different spins can be really ignored in generic situations, where the electron-electron interaction is irrespective of spin. 
We seek to find an answer to this question by investigating what happens in a fractionally filled, lowest Landau level with spin-dependent holomorphicity, i.e., electrons of one spin species reside in the holomorphic lowest Landau level, while those of the other in the antiholomorphic counterpart.  
If the correlation between electrons with different spins can be really ignored, the ground state will be given by a simple product of two independent copies of the FQHS, which can be mapped onto the independent bipartite FTI in the lattice via the mapping of the basis functions mentioned above.

To test this scenario, we construct and perform exact diagonalization of an appropriate model Hamiltonian realizing spin-dependent holomorphicity. 
As a result, here, we show that generic, strongly correlated, fractionally filled states cannot be described as two independent copies of the FQHS, suggesting that FTIs in the lattice cannot be described as those of the FCI either.
In this system, a fractionally filled topological insulator is predicted to occur exclusively at half filling, susceptible to an inherent spontaneous symmetry breaking, leading to the spatial separation of spins.
This half-filled spin-holomorphic FTI and its lattice version can be used as a {\it spin filter,} which could be potentially useful in spintronics~\cite{Zutic04}.

%%%%%%%
%%%%%%%
{\bf Results}
%%%%%%%
%%%%%%%

%%%%%%%%%%%%%%%%%%%%
{\bf Spin-holomorphic Landau levels.}
%%%%%%%%%%%%%%%%%%%%
We begin by constructing an appropriate model Hamiltonian realizing spin-dependent holomorphicity. 
Actually, Bernevig and Zhang~\cite{Bernevig06_PRL} proposed essentially the identical model Hamiltonian to describe the dynamics of electrons confined in a two-dimensional parabolic quantum well with effective spin-orbit coupling induced by an appropriate strain gradient. 
In this work, we consider a somewhat different physical mechanism for the realization of such model Hamiltonian.

%%%%%%%%%%%%%%%%%%%%%%%%%%%%%%%%%%%%%%%%%%%%%%%%%%%%%%%%%%%%%%%%%%%%%%%%%%%%%%
\begin{figure*}
\begin{center}
\includegraphics[width=2\columnwidth,angle=0]{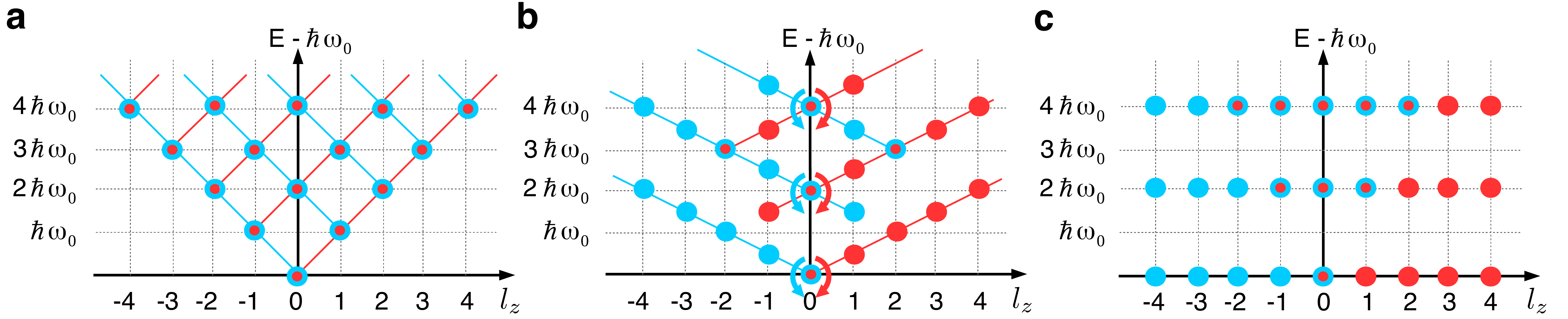}
\caption{
{\bf Formation of the spin-holomorphic Landau levels.}
({\bf a}) The red and blue (mixed) circles denote the (coincidental) energy levels of spin up and down electrons, respectively, in a two-dimensional harmonic oscillator as a function of the $z$-component angular momentum eigenvalue $l_z$.
The energy levels are independent of spin in the absence of spin-orbit coupling.
({\bf b}) With addition of spin-orbit coupling, the energy levels of spin up and down electrons evolve differently.
The red and blue lines are a guide to eye, showing that the energy levels threaded by the respective lines are rotated around the $l_z=0$ circles to the opposite directions, as indicated by the corresponding arrows. 
Note that the sign of the spin-orbit coupling constant is not important since different signs just interchange the role of spin up and down electrons. 
({\bf c}) At an appropriate value of the spin-orbit coupling constant, the energy levels form effective Landau levels with the effective magnetic fields being opposite for different spins. 
We call these lowest effective Landau levels the spin-holomorphic Landau levels since their holomorphicity depends on spin. 
}
\label{fig:Spin-holomorphic_LL}
\end{center}
\end{figure*}
%%%%%%%%%%%%%%%%%%%%%%%%%%%%%%%%%%%%%%%%%%%%%%%%%%%%%%%%%%%%%%%%%%%%%%%%%%%%%%

Fundamentally, any spin-orbit coupling owes its origin to the relativistic nature of the Dirac Hamiltonian.
Specifically, the usual ${\bf L}\cdot{\bf s}$ term can be obtained from the $({\bf p}\times{\bf E})\cdot{\bf s}$ term, which is generated by expanding the Dirac Hamiltonian in the non-relativistic limit.
If the electric field ${\bf E}$ is induced by a radial electrostatic potential $V(r)$, i.e., ${\bf E}=-\frac{1}{r}\frac{d V}{d r}$ {\bf r}, the Dirac Hamiltonian can be expanded in the non-relativistic limit as follows:
\begin{align}
\label{eq:Hamiltonian_radial}
H= \frac{{\bf p}^2}{2m_e} +e V(r) +{\cal C} \frac{1}{r} \frac{d V(r)}{d r} {\bf L}\cdot\boldsymbol{\sigma},
\end{align}
where ${\cal C}$ is equal to $e/4m_e^2 c^2$ in vacuum, but is assumed to be varied in material. 
Similarly, the electron mass $m_e$ can be also varied from its vacuum value in material.
If $V(r)$ is a two-dimensional parabolic potential confined in the $xy$ plane, Eq.~\eqref{eq:Hamiltonian_radial} can be further simplified as follows:
\begin{align}
\label{eq:Hamiltonian_parabolic}
H= \frac{{\bf p}^2}{2m_e} +\frac{1}{2} m_e \omega_0^2 r^2 -\alpha L_z \sigma_z,
\end{align}
where the spin-orbit coupling constant $\alpha$ ($=-{\cal C} m_e \omega_0^2$) is independent of $r$. 
Note that ${\bf L}\cdot\boldsymbol{\sigma}$ reduces to $L_z \sigma_z$ due to the two-dimensional confinement, which also requires that ${\bf p}^2=p_x^2+p_y^2$ and $r^2=x^2+y^2$.

There is a close similarity between Eq.~\eqref{eq:Hamiltonian_parabolic} and the Hamiltonian for the Landau levels in the circular gauge, ${\bf A}=\frac{B}{2}\hat{z}\times{\bf r}$:
\begin{align}
\label{eq:Hamiltonian_circular_gauge}
H= \frac{{\bf p}^2}{2m_e} +\frac{1}{2} m_e \left( \frac{\omega_c}{2} \right)^2 r^2 -\frac{\omega_c}{2} L_z ,
\end{align}
where the cyclotron frequency $\omega_c=eB/m_e c$.
When $\alpha = \omega_0$, the Hamiltonian in Eq.~\eqref{eq:Hamiltonian_parabolic} generates exactly the same Landau levels as that in Eq.~\eqref{eq:Hamiltonian_circular_gauge} (with the level spacing replaced by $2 \hbar \omega_0$ and the relevant length scale by $l_0=\sqrt{\hbar /2 m_e \omega_0}$) except for a salient distinction that electrons with different spins now feel opposite effective magnetic fields.  
In particular, the holomorphicity of the lowest effective Landau levels depends on spin, i.e., 
$\phi_{m\uparrow}({\bf r})=\langle {\bf r}|c^\dagger_{m\uparrow}|0\rangle \propto z^m e^{-zz^*/4l_0^2}$ and $\phi_{m\downarrow}({\bf r})=\langle {\bf r}|\bar{c}^\dagger_{m\downarrow}|0\rangle \propto (z^*)^m e^{-zz^*/4l_0^2}$, where $c^\dagger_{m\uparrow}$ and $\bar{c}^\dagger_{m\downarrow}$ are the respective creation operators for spin up and down electrons in the holomorphic and antiholomorphic orbitals with quantum number $m$.
For later convenience, we set $m=l_z$ and $-l_z$ for the holomorphic and antiholomorphic orbitals, respectively, with $l_z$ being the actual $z$-component angular momentum eigenvalue.
For clarity, the antiholomorphic creation operators are distinguished from the holomorphic counterparts via the bar on top.
Let us call these lowest effective Landau levels the spin-holomorphic Landau levels. 
At general values of $\alpha$, the Hamiltonian in Eq.~\eqref{eq:Hamiltonian_parabolic} can be regarded as two copies of the Fock-Darwin Hamiltonian with opposite magnetic fields for different spins, whose energy eigenvalues are given by $E=\hbar\omega_0 (2n+1 \pm l_z) \mp \hbar\alpha l_z$ with $n=0, 1, 2, \cdots$ and $l_z=\mp n, \mp n \pm 1, \mp n \pm 2, \cdots$ for spin up and down electrons, respectively.
See Fig.~\ref{fig:Spin-holomorphic_LL} for illustration.

To investigate the effects of correlation, which is responsible for the very emergence of FTIs, we next consider the Coulomb interaction between electrons in the spin-holomorphic Landau levels. 
To this end, we have to choose appropriate geometries for the system.

One of the most convenient geometries is the spherical geometry, where the system is placed on the the surface of a sphere with a Dirac monopole located at the center~\cite{Wu76,Haldane83,Jain_Book}. 
Mathematically, all the basis functions in the planar geometry can be one-to-one mapped to those in the spherical geometry via the stereographic projection, i.e., the $L_z$ eigenstates in the planar geometry with $l_z=\{0,\cdots,M\}$ and $\{0,\cdots,-M\}$ are mapped to those in the spherical geometry with $l_z=\{M/2,\cdots,-M/2\}$ and $\{-M/2,\cdots,M/2\}$ for a positive and negative magnetic field, respectively, where $M$ is the maximum absolute value of $l_z$ defining the system size and thus the filling factor.
In the spin-holomorphic situation, we consider a spin-dependent Dirac monopole. 
Specifically, the spin-dependent monopole strength $2Q_{\uparrow/\downarrow}$ is related to the spin up/down electron number $N_{\uparrow/\downarrow}$ via $2Q_{\uparrow/\downarrow}=\pm(\nu_{\uparrow/\downarrow}^{-1} N_{\uparrow/\downarrow} -\lambda)$, where $\nu_{\uparrow/\downarrow}$ is the spin up/down filling factor and $\lambda$ is the so-called flux shift.
Let us call this geometry the spin-holomorphic spherical geometry. 
In this work, we focus on the time-reversal symmetric situation with $N_\uparrow=N_\downarrow$ and $\nu_\uparrow=\nu_\downarrow$.
%Also, considering the spin degree of freedom, half filling is defined such that $\nu_\uparrow=\nu_\downarrow=1/2$, and therefore $\nu_{\rm tot}=\nu_\uparrow+\nu_\downarrow=1$, which is similar to the definition of half filling in the Hubbard model. 
%Other fillings are defined similarly. 

%%%%%%%%%%%%%%%%%%%%%%%%%%%%%%%%%%%%%%%%%%%%%%%%%%%%%%%
\begin{figure}
\includegraphics[width=\columnwidth,angle=0]{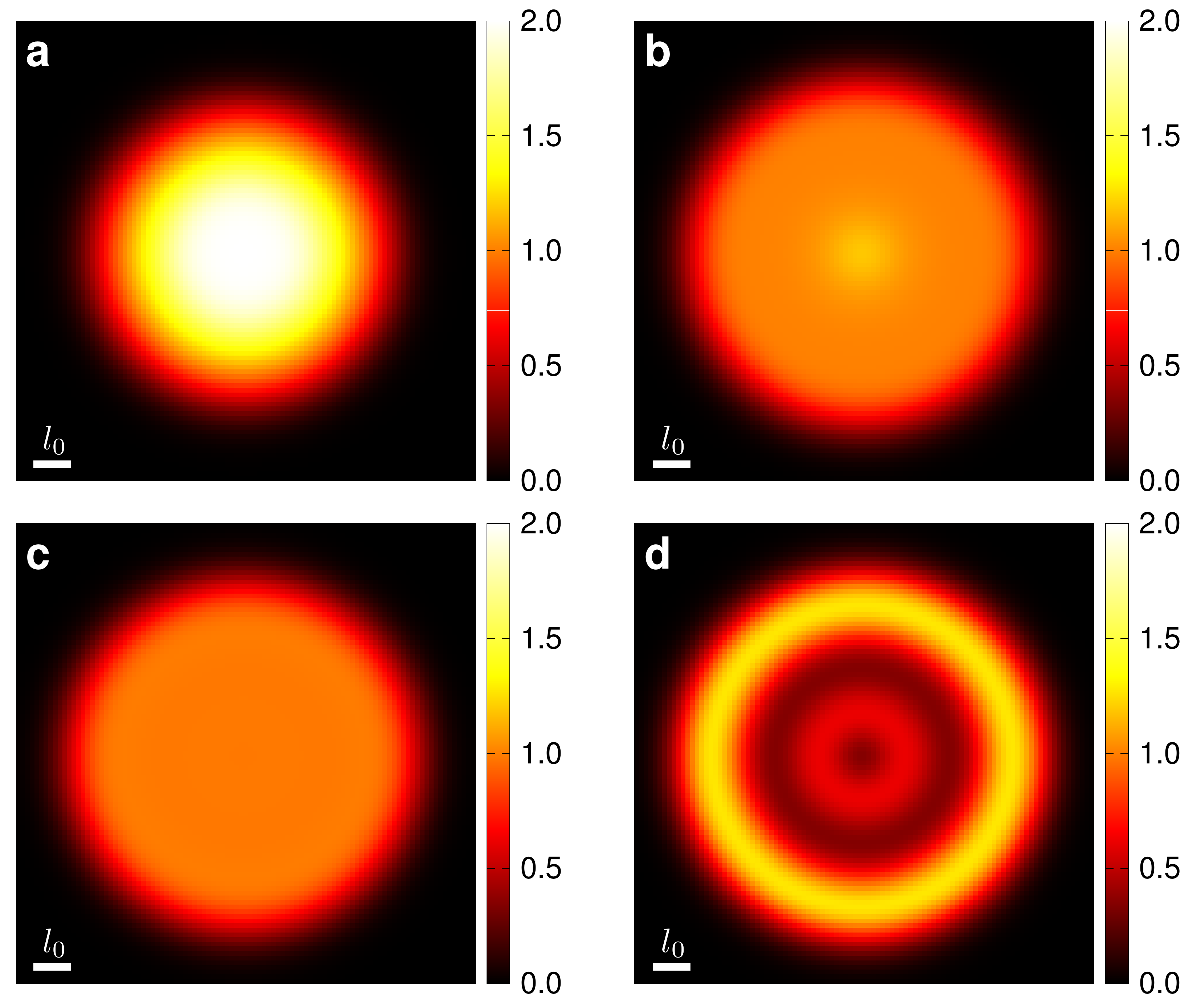}
\caption{
{\bf Electron density profile as a function of residual confining potential strength $\gamma$ at half filling in the spin-holomorphic disc geometry.}
$\gamma=\omega_0-\alpha$ with $\omega_0$ being the natural frequency of the parabolic potential well and $\alpha$ being the spin-orbit coupling constant.
({\bf a}) $\gamma=0.25$, ({\bf b}) 0.125, ({\bf c}) 0.105, and ({\bf d}) 0.0 in units of $e^2 /\epsilon l_0 $, where $l_0=\sqrt{\hbar /2 m_e \omega_0}$ is the natural length scale of the system.
Scale bars in the figure denote $l_0$, showing the overall size of the electron droplet.
Also, $N=N_\uparrow+N_\downarrow=16$ and $m_{\rm max} = 15$.
}
\label{fig:Disc_density_profile}  
\end{figure}   
%%%%%%%%%%%%%%%%%%%%%%%%%%%%%%%%%%%%%%%%%%%%%%%%%%%%%%%%%%   

Another convenient geometry is the planar geometry as described in Eq.~\eqref{eq:Hamiltonian_parabolic} with an appropriate boundary condition, without which electrons would spread out unboundedly.
One way to prevent the spreading of an electron droplet is to introduce an additional confining potential. 
Fortunately, there is a natural confining potential present in the system; when the spin-orbit coupling constant $\alpha$ is not exactly equal to $\omega_0$, there is a residual confining potential $\gamma L_z \sigma_z$ with $\gamma=\omega_0-\alpha$.  
If $\gamma$ gets too large, electrons are all squeezed tightly into the center.
On the other hand, if $\gamma$ gets too small, the electron droplet falls apart completely, and electrons are all pushed to an artificial system boundary at $r \simeq \sqrt{2m_{\rm max}} l_0$, where $m_{\rm max}$ is a preset maximum value of $|l_z|$.
In this situation, the electron density profile would be of ring shape, whose radius increases as a function of $m_{\rm max}$.
Only if $\gamma$ lies within a right range, the electron droplet can have a roughly uniform electron density in a natural disc area, which is independent of $m_{\rm max}$.
Let us call this geometry the spin-holomorphic disc geometry.
We find that, for $N = N_\uparrow+N_\downarrow \sim 10 \mbox{--} 16$ at half filling, $\gamma \simeq 0.1 e^2/\epsilon l_0$ gives rise to a healthy competition between the residual confining potential and the Coulomb interaction energies so that the electron density is maintained to be unity (i.e., $\nu_{\rm tot}=1$) more or less uniformly throughout the entire electron droplet, which does not change much for any $m_{\rm max}$ larger than $N-1$. 
Figure~\ref{fig:Disc_density_profile} shows the evolution of the electron density profile as a function of $\gamma$.

In what follows, we perform exact diagonalization of the Coulomb interaction Hamiltonian in both spin-holomorphic spherical and disc geometries.
Specifically, we diagonalize the following Hamiltonian in the spin-holomorphic spherical geometry:
\begin{align}
H={\cal P}_{\rm SHLL} \left( \frac{e^2}{\epsilon l_0} \sum_{i<j} \frac{1}{r_{ij}} \right) {\cal P}_{\rm SHLL},
\end{align}
where ${\cal P}_{\rm SHLL}$ is the projection operator onto the spin-holomorphic Landau levels.
In the spin-holomorphic disc geometry, there is an additional term due to the residual confining potential: 
$H^\prime=\gamma \sum_i L_{z,i} \sigma_{z,i}$.
See {\bf Methods} for the details of the Coulomb matrix elements in both spin-holomorphic geometries.

\begin{figure}
\includegraphics[width=\columnwidth,angle=0]{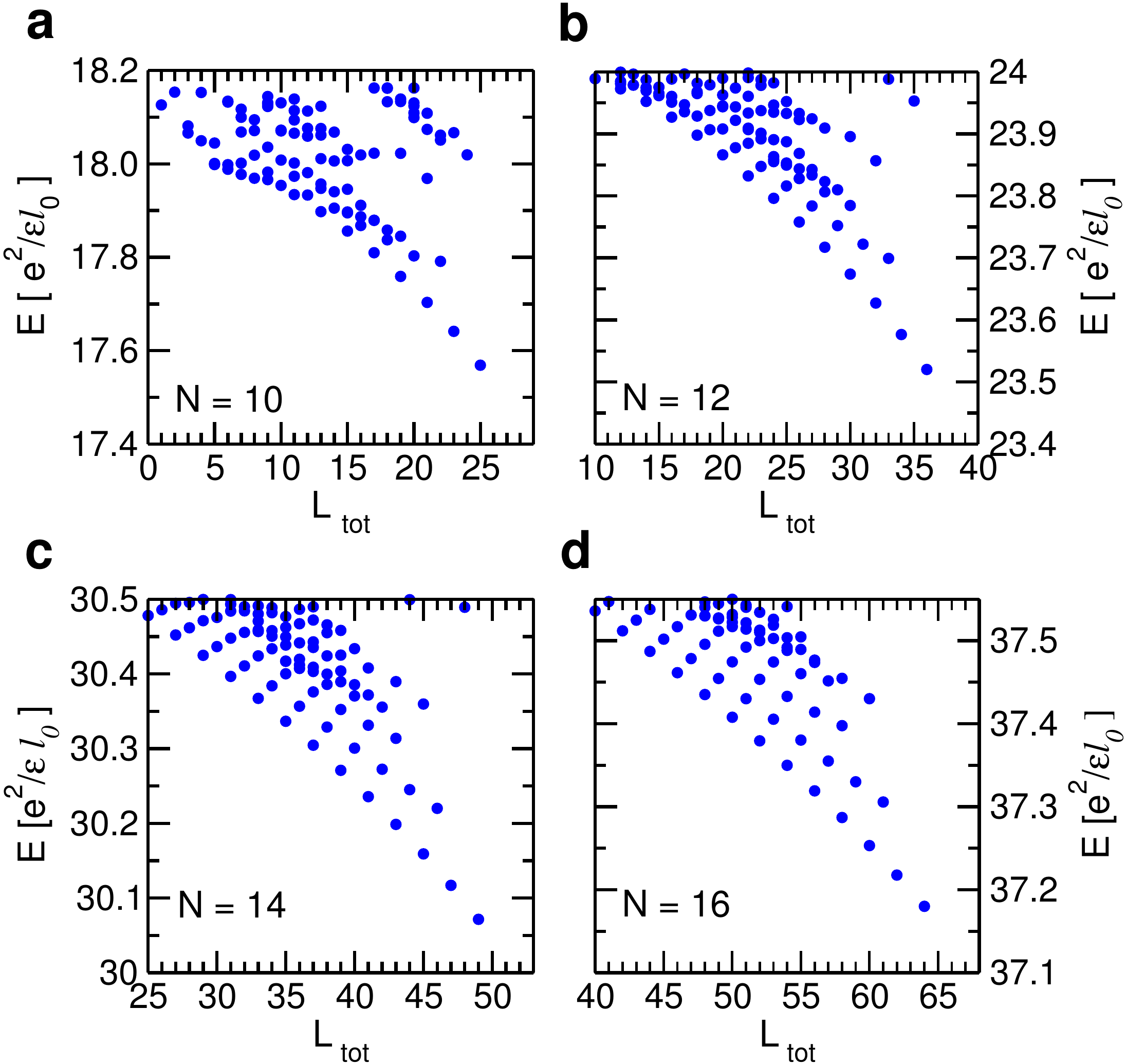}
\caption{
{\bf Exact energy spectra as a function of total angular momentum quantum number $L_{\rm tot}$ at half filling in the spin-holomorphic spherical geometry.}
The particle number $N$ is varied with ({\bf a}) $N=10$, ({\bf b}) 12, ({\bf c}) 14, and ({\bf d}) 16.
Here, the $z$-component total angular momentum quantum number, $L_{{\rm tot},z}$, is set equal to zero.
Similar to the usual spherical geometry, each state at a given value of $L_{\rm tot}$ has $2L_{\rm tot}+1$ degenerate energy partners with $L_{{\rm tot},z}$ ranging from $-L_{\rm tot}$ to $L_{\rm tot}$. 
}
\label{fig:Energy_spectra}
\end{figure}

%%%%%%%%%%%%%%%%%%%%%%%%%%%%%%
{\bf Spontaneous symmetry breaking and spin separation.}
%%%%%%%%%%%%%%%%%%%%%%%%%%%%%%
We first show exact diagonalization results obtained in the spin-holomorphic spherical geometry, which, being edgeless, can reveal bulk properties more clearly. 
Specifically, Figure~\ref{fig:Energy_spectra} shows the exact energy spectra as a function of total angular momentum quantum number $L_{\rm tot}$ at half filling in the spin-holomorphic spherical geometry.

Before discussing the physical meaning of the results in detail, it is important to note that the total angular momentum operator, ${\bf L}_{\rm tot}=\sum_{i=1}^N {\bf L}_i$, should be appropriately generalized to take care of spin-dependent holomorphicity so that the eigenvalue of ${\bf L}_{\rm tot}^2$ remains as a good quantum number, being $L_{\rm tot}(L_{\rm tot}+1)$ as usual. 
See {\bf Methods} for details.
Also, considering the spin degree of freedom, half filling is defined such that $\nu_\uparrow=\nu_\downarrow=1/2$, and therefore $\nu_{\rm tot}=\nu_\uparrow+\nu_\downarrow=1$, which is similar to the definition of half filling in the Hubbard model. 
Specifically, at half filling, $2Q_{\uparrow/\downarrow}=\pm(2N_{\uparrow/\downarrow}-1)$, or $N=2|Q_{\uparrow/\downarrow}|+1$, meaning that the total number of electrons is exactly the same as the available orbitals in each of the holomorphic and antiholomorphic Landau levels.
Unless stated otherwise, we focus on half filling since other fractionally filled states are generally compressible, as explained in the next section.

%%%%%%%%%%%%%%%%%%%%%%%%%%%%%%%%%%%%%%%%%%%%%%%%%%%%%%%%%%%%%%%%%%%%%%
%%%%%%%%%%%%%%%%%%%%%%%%%%%%%%%%%%%%%%%%%%%%%%%%%%%%%%%%%%%%%%%%%%%%%%

As shown in Fig.~\ref{fig:Energy_spectra}, the behavior of the exact energy spectra in the half-filled spin-holomorphic Landau levels is in stark contrast to that of the usual FQHSs, exhibiting a remarkable feature that the ground state occurs at the maximum value of total angular momentum quantum number, i.e., $L_{\rm tot}=L_{\rm tot}^{\rm max}=N^2/4$. 
Usually, the FQHSs occur at $L_{\rm tot}=0$ with a well-developed energy gap, indicating that they are uniform, incompressible states.
Even when the system becomes compressible, the ground state is supposed to occur at a random, but not the maximum value of $L_{\rm tot}$. 
In the current situation, the ground state occurs consistently at $L_{\rm tot}=L_{\rm tot}^{\rm max}$, which diverges even in the proper scaling as the system size grows, i.e., $L_{\rm tot}^{\rm max}/\sqrt{|Q_{\uparrow/\downarrow}|} \sim N^{3/2}$. 
This does not only mean that the ground state is non-uniform, but also that there would be an infinite number of degenerate states in the thermodynamic limit. 
What would this mean physically?

To understand the physical meaning of this result, it is important to realize that there are certain states among the above vastly degenerate multiplets at $L_{\rm tot}=L_{\rm tot}^{\rm max}$, whose exact wave functions are uniquely determined by the symmetry alone.  
Such a state is the state at $L_{{\rm tot},z}=L_{\rm tot}^{\rm max}$, whose exact wave function is obtained uniquely by filling all the orbitals in the upper and lower hemispheres with spin up and down electrons, respectively.
The other is the state at $L_{{\rm tot},z}=-L_{\rm tot}^{\rm max}$, whose exact wave function is obtained similarly with the roles of spin up and down electrons interchanged.
This means that, in these states, different spins are spatially separated from each other completely.

The spin separation in the states at $L_{{\rm tot},z}=\pm L_{\rm tot}^{\rm max}$ can be shown more clearly by examining the explicit forms of their exact wave functions:
\begin{align}
\Psi_{\pm L_{\rm tot}^{\rm max}}= \prod_{m=1}^{N/2} c^\dagger_{\pm m\uparrow} \bar{c}^\dagger_{\mp m\downarrow} |0\rangle ,
\end{align}
%where $c^\dagger_{m\uparrow}$ and $\bar{c}^\dagger_{m\downarrow}$ are the respective creation operators for spin up and down electrons in the holomorphic and antiholomorphic orbitals with quantum number $m$.
%For clarity, the antiholomorphic creation operators are distinguished from the holomorphic counterparts in terms of the bar on top.  
where $c^\dagger_{m\uparrow}$ and $\bar{c}^\dagger_{m\downarrow}$ are the respective creation operators for spin up and down electrons in the spin-holomorphic spherical geometry.
More intuitively, $\Psi_{\pm L_{\rm tot}^{\rm max}}$ can be denoted as $|\uparrow,\cdots,\uparrow,\downarrow\cdots,\downarrow\rangle$ and $|\downarrow,\cdots,\downarrow,\uparrow,\cdots,\uparrow\rangle$, respectively, clearly showing the spatial separation of spins.

The above wave function is definitely not of the independent bipartite form, which is to be given by the direct product of two composite-fermion (CF) seas at half filling with complex conjugation applied to the spin-down part. i.e., $\Psi_{\rm ind. bip.}=\Psi_{^2{\rm CFS}_\uparrow} \otimes \Psi^*_{^2{\rm CFS}_\downarrow}$, where $^2$CFS stands for the Fermi sea of CFs carrying two vortices~\cite{Jain_Book,Jain89,Kalmeyer92,Halperin93}. 
Therefore, it is shown that the correlation between electrons with different spins definitely cannot be ignored, at least for the states at $L_{{\rm tot},z}=\pm L_{\rm tot}^{\rm max}$.
As shown below, the same is true for the other degenerate multiplets.
%As shown below, the same is true for other general fillings, say 1/3 filling.

Meanwhile, there is a close similarity between the usual bilayer quantum Hall and the current spin-holomorphic systems, suggesting that a natural trial state could be given by the spin-holomorphic version of the Halperin (111) state~\cite{Halperin83,Park04,Park06}:
\begin{align}
\Psi_{(1\bar{1}1)} = \prod_{m=-N/2}^{N/2} (c^\dagger_{m\uparrow}+\bar{c}^\dagger_{m\downarrow}) |0\rangle ,
\end{align}
which, in terms of the real space coordinates, corresponds to $\prod_{i<j} (z_i-z_j) \prod_{m<n} (\omega^*_m-\omega^*_n) \prod_{k,l}(z_k-\omega^*_l)$ with $z$ and $\omega$ denoting the coordinates of the spin up and down electrons, respectively. 
Bernevig and Zhang~\cite{Bernevig06_PRL} previously used a similar logic to propose the spin-holomorphic version of the Halperin $(mmn)$ state as the trial state for a generic FTI.
It is interesting to note that $\Psi_{(1\bar{1}1)}$ contains $\Psi_{\pm L_{\rm tot}^{\rm max}}$ as a subset of the constituent states participating in the linear superposition. 
In some sense, $\Psi_{(1\bar{1}1)}$ is analogous to the paramagnetic state, which, in addition to prevalent random fluctuations, contains various ferromagnetic constituent states, to which $\Psi_{\pm L_{\rm tot}^{\rm max}}$ are analogous, spontaneously breaking the spatial symmetry. 
We provide a more detailed explanation for this analogy below.

Other degenerate multiplets at $L_{{\rm tot},z} \neq \pm L_{\rm tot}^{\rm max}$ have generally very complicated wave functions, whose amplitudes are spread over various many-body bases in a seemingly uncoordinated fashion. 
It is, however, important to realize that all these states can be uniquely obtained by applying the angular momentum lowering and raising operator to the states at $L_{{\rm tot},z}=\pm L_{\rm tot}^{\rm max}$.  
That is, they are all related with each other via rigid rotation, which, combined with the very existence of degeneracy, means that these states can be linearly re-superposed to produce other spin-separated states just like those at $L_{\rm tot}=\pm L_{\rm tot}^{\rm max}$ with their spin-separation lines being different from the equator.
In other words, the spin-separation line can be freely rotated to become any of the great circles. 
Eventually, such a freedom manifests itself as a spontaneous breaking of the spatial symmetry.

%%%%%%%%%%%%%%%%%%%%%%%%%%%%%%%%%%%%%%%%%%%%%%%%%%%%%%%%%%%%%%%%%
\begin{figure}
\includegraphics[width=\columnwidth,angle=0]{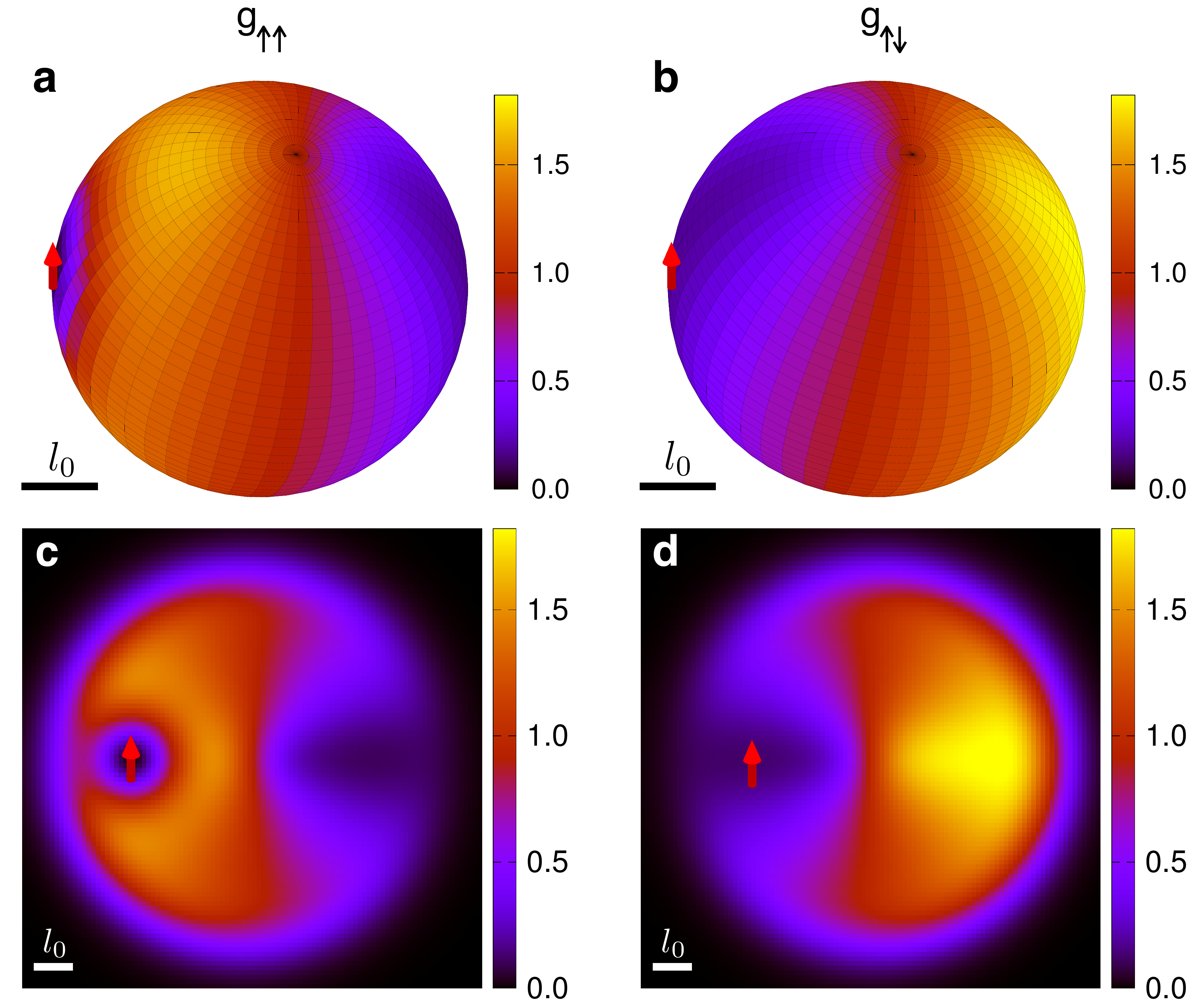}
\caption{
{\bf Pair correlation functions in the half-filled spin-holomorphic Landau levels.}
Pair correlation functions between the same spin, ({\bf a}) $g_{\uparrow\uparrow}(\bold r)$, and between different spins, ({\bf b}) $g_{\uparrow\downarrow}(\bold r)$ in the spin-holomorphic spherical geometry for the state at $L_{\rm tot}=64$ and $L_{{\rm tot}, z}=0$ with $N=16$.
Note that the reference electron is chosen to be spin up and located on the equator, as indicated by the red arrow in the figure.
Similar pair correlation functions, ({\bf c}) $g_{\uparrow\uparrow}(\bold r)$ and ({\bf d}) $g_{\uparrow\downarrow}(\bold r)$, are shown in the spin-holomorphic disc geometry with $N=16$.
Here, the residual confining potential strength is set to be $\gamma=0.105 e^2/\epsilon l_0$ to ensure a uniform electron density.
}
\label{fig:Pair_correlation}  
\end{figure}   
%%%%%%%%%%%%%%%%%%%%%%%%%%%%%%%%%%%%%%%%%%%%%%%%%%%%%%%%%%%%%%%%%%%%   

To provide evidence for this interpretation, we compute the pair correlation function, which measures the probability of finding an electron at position ${\bf r}$ when a reference electron is located at the origin. 
Among the various degenerate multiplets at $L_{\rm tot}=L_{\rm tot}^{\rm max}$, we focus on the state at  $L_{{\rm tot},z}=0$, which is expected to be a linearly superposed state of all possible spin-separated states, whose spin-separation lines are the great circles connecting the north and south poles. 
In this situation, locating the origin on the equator would select a particular spin-separated constituent state, pinning the spin-separation line. 
For this selected spin-separated state, the pair correlation function between the same spin should be small at long distance, while large at short distance except for the obvious drop at the origin due to the formation of an exchange hole required by the Pauli exclusion principle.
The pair correlation function between different spins should show the opposite behavior.
Figure~\ref{fig:Pair_correlation}~(a) and (b) show that this is indeed exactly the case, confirming the spin separation.
%It is important to note that the spin separation emerges via a spontaneous breaking of the spatial symmetry.

Next, we check if such a spin separation also occurs in the spin-holomorphic disc geometry.
Figure~\ref{fig:Pair_correlation}~(c) and (d) show the pair correlation functions between the same and different spins, respectively, in the spin-holomorphic disc geometry. 
As one can see, the behaviors of the pair correlation functions are exactly as expected for the spin-separated state, consistent with the results obtained in the spin-holomorphic spherical geometry.

Fundamentally, the spin separation is due to a peculiar structure of the Coulomb matrix elements in the spin-holomorphic Landau levels.
In the usual quantum Hall system, electrons are always scattered away from each other regardless of spin. 
In the spin-holomorphic system, however, the angular momentum conservation law dictates that, after scattering, electrons with different spins move together to the same radial direction, making it difficult to reduce the Coulomb energy unless electrons with different spins are spatially separated from the outset and thus have no chance to encounter each other. 
See {\bf Methods} for details.

So far, it has been shown that different spins are spatially separated in the half-filled spin-holomorphic Landau levels.  
In what follows, we show that such a spin-separated state should be incompressible.

%%%%%%%%%%%%%%%%%%%%%%%%%%%%%%%%
{\bf Energy gap, helical edge excitation, and incompressibility.}
%%%%%%%%%%%%%%%%%%%%%%%%%%%%%%%%
At half filling, the total number of electrons is exactly the same as the available orbitals in each of the holomorphic and antiholomorphic Landau levels.
Meanwhile, it is shown above that different spins are spatially separated in the half-filled spin-holomorphic Landau levels.
We argue that these two facts lead to incompressibility.

The logic is as follows.
The combination of the above two facts make all the orbitals in the half-filled spin-holomorphic Landau levels occupied by either spin up or down electrons without any vacancy.  
Then, any additional electron of a given spin species cannot penetrate into the region of the same spin due to the Pauli exclusion principle and thus be pushed to the region of the opposite spin.
However, there would be a large Coulomb energy cost for this to happen due to the spin separation. 
Therefore, the spin-separated half-filled state should be incompressible.

The same logic actually tells us that fractionally filled states at less than half filling should be generally compressible since there are now vacant orbitals between two spin-separated regions. 
In this situation, any additional electron can nestle nicely into these vacant orbitals without costing too much Coulomb energy. 
%disturbing the stability of the spin-separated state itself.
Fractionally filled states at greater than half filling are also expected to be compressible owing to the particle-hole symmetry.

%%%%%%%%%%%%%%%%%%%%%%%%%%%%%%%%%%%%%%%%%%%%%%%%%%%%%%%%%%%
\begin{figure}
\includegraphics[width=\columnwidth,angle=0]{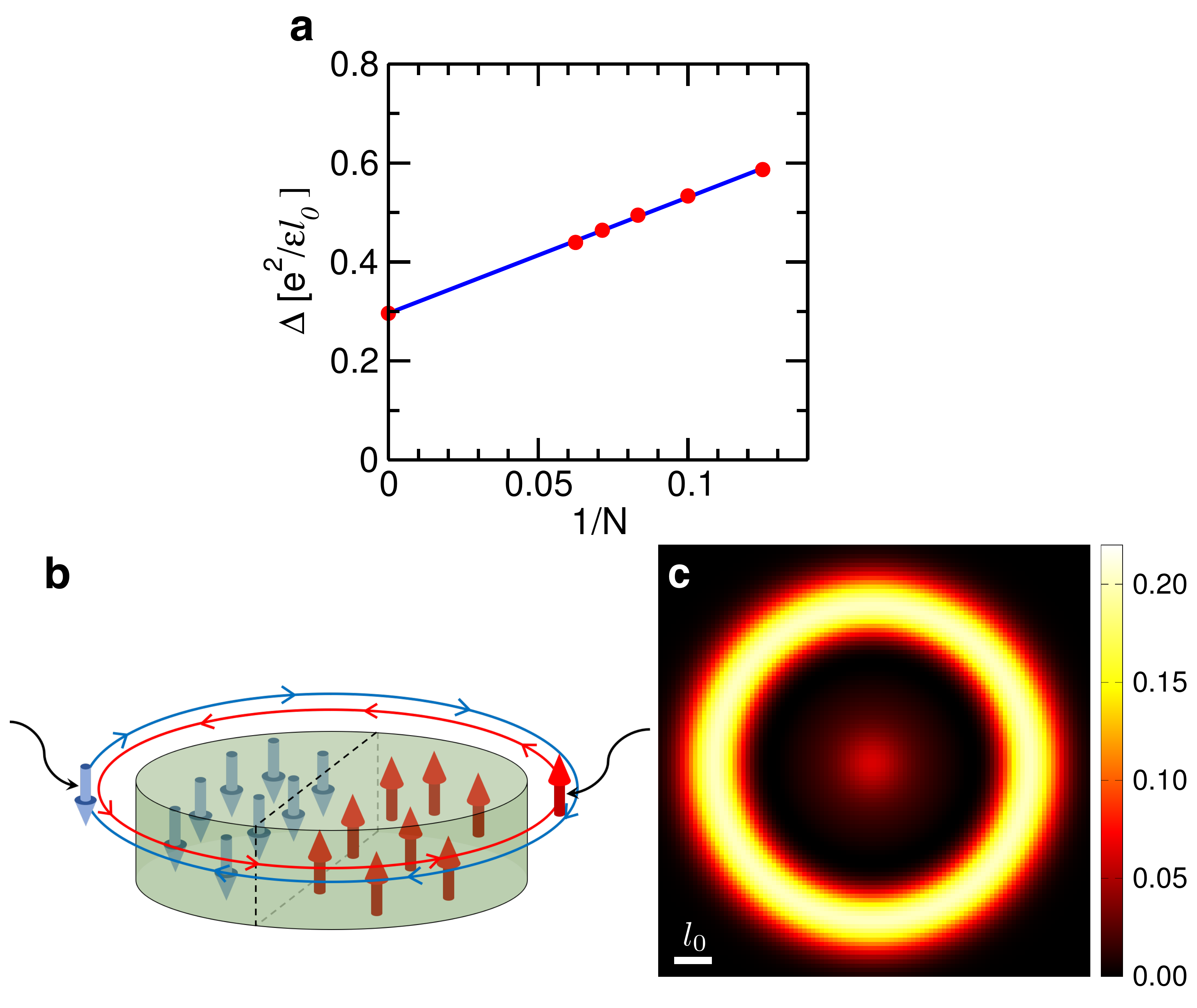}
\caption{
{\bf Incompressibility of the half-filled spin-holomorphic fractional topological insulator.}
({\bf a}) Energy gap as a function of inverse particle number $1/N$ in the spin-holomorphic spherical geometry.
({\bf b}) Schematic diagram showing the helical edge excitations in the spin-holomorphic disc geometry. 
({\bf c}) Difference of the electron density profiles between before and after adding two extra electrons of up and down spins at the system with $N=14$. 
Here, the residual confining potential strength $\gamma$ is set to be $0.105 e^2/\epsilon l_0$.
}
\label{fig:Incompressibility}
\end{figure}
%%%%%%%%%%%%%%%%%%%%%%%%%%%%%%%%%%%%%%%%%%%%%%%%%%%%%%%%%%%

To be concrete, we compute the energy gap of the half-filled state in the spin-holomorphic spherical geometry by using the following formula:
\begin{align}
\Delta= E_{N+1,Q_{\uparrow/\downarrow}}+E_{N-1,Q_{\uparrow/\downarrow}}-2E_{N,Q_{\uparrow/\downarrow}},
\end{align}
where $E_{N,Q_{\uparrow/\downarrow}}$ is the ground state energy of $N$ particles at flux $Q_{\uparrow/\downarrow}$.
Above, we increase and decrease the particle number by one from $N$ satisfying the half-filling condition, $N=2|Q_{\uparrow/\downarrow}|+1$.
It is important to note that the energy gap is given as the sum of $E_{N+1,Q_{\uparrow/\downarrow}}-E_{N,Q_{\uparrow/\downarrow}}$ and $E_{N-1,Q_{\uparrow/\downarrow}}-E_{N,Q_{\uparrow/\downarrow}}$ to take into account the chemical potential shift associated with the particle number change at a fixed $Q_{\uparrow/\downarrow}$. 
Figure~\ref{fig:Incompressibility}~(a) shows that, plotted as a function of $1/N$, the energy gap can be nicely linearly extrapolated to a finite value $\sim 0.3 e^2/\epsilon l_0$ in the thermodynamic limit, confirming the incompressibility of the half-filled spin-holomorphic state, which can be now regarded as a legitimate fractional topological {\it insulator}.

Meanwhile, we have confirmed that a fractionally-filled spin-holomorphic state at less than half filling, say, $\nu_\uparrow=\nu_\downarrow=1/3$, has a vanishing energy gap in the thermodynamic limit, which is consistent with the previous result reported by Chen and Yang~\cite{Chen12} that sufficiently strong interspin interaction generates a compressible state with phase separation at this filling.

In the disc geometry, incompressibility means the absence of low-energy excitations in the bulk.
If so, any additional electron would be pushed to the edge, creating an edge excitation. 
In the presence of spin-dependent holomorphicity, the edge excitations with different spins should form mutually counter-rotating orbitals on the edge, i.e., helical edge excitations. 
See Fig.~\ref{fig:Incompressibility}~(b) for a schematic diagram. 
To confirm this scenario, we compute the difference of the electron density profiles between before and after adding two extra electrons of up and down spins in the half-filled spin-holomorphic state.
Figure~\ref{fig:Incompressibility}~(c) shows that almost all of the probability weights of the two extra electrons are indeed pushed to the edge, forming the helical edge excitations.

Now, let us discuss the transport signature of such helical edge excitations.
In the absence of magnetic impurities, the system is time-reversal symmetric, and thus any back-scattering, or hybridization between helical edge excitations with different spins is completely prohibited, protecting the Kramer's degeneracy.
Then, the Landauer-B\"{u}ttiker formalism \cite{Buttiker88} for the ballistic transport suggests that the Hall conductivity of each spin species should be quantized exactly as $\pm e^2/h$. 
This means that the usual Hall conductivity is zero, while the spin Hall conductivity is quantized as $2 e^2/h$ just like in the usual 2D topological insulators~\cite{Kane05_PRL_Quantum,Bernevig06_Science,Konig07}. 
It is important to note that, even though the spin Hall conductivity is $2 e^2/h$, the total filling factor $\nu_{\rm tot}=\nu_\uparrow+\nu_\downarrow=1$, not 2 as in the usual topological insulators, resulting in an intriguing mismatch between the spin Hall conductivity in units of $e^2/h$ and the total filling factor.

%%%%%%%%%
%%%%%%%%%
{\bf Discussion}
%%%%%%%%%
%%%%%%%%%

In this work, we provide evidence that generic, strongly correlated states in a  fractionally filled, topologically non-trivial band cannot be described as two independent copies of the FQHS or FCI.
Specifically, we construct and perform exact diagonalization of an appropriate model Hamiltonian, which forms effective Landau levels with spin-dependent holomorphicity, i.e., electrons with different spins experience opposite effective magnetic fields.

It is predicted that the FTI occurring at half filling in the spin-holomorphic Landau levels is susceptible to an inherent spontaneous symmetry breaking, leading to the spatial separation of spins. 
The half-filled spin-holomorphic FTI with such spin separation could be potentially useful in spintronics since it can serve as a robust interaction-driven {\it spin filter,} which sorts electrons with different spins into two spatially separated regions. 
Once embedded in their respective spin-separated regions, spins are expected to be protected against various decoherence mechanisms by the Coulomb interaction. 
A spin-filtered current can flow by attaching a lead deep inside the desired spin-separated region, where the edge current surrounding the attached lead should be composed of only the single spin species corresponding to the region.

Now, we discuss briefly how this state can be realized in experiments. 
Our model Hamiltonian is based on a two-dimensional electron gas confined in a parabolic confining potential with strong spin-orbit coupling, which can be realized by constructing a semiconductor quantum well or heterostructure on the substrate made of a strong spin-orbit-coupled material. 
Another way of realizing essentially the same model Hamiltonian is to induce an effective spin-orbit coupling by applying an appropriate strain gradient in a two-dimensional parabolic quantum well, as proposed by Bernevig and Zhang~\cite{Bernevig06_PRL}.

Perhaps, a more important possibility is the half-filled Chern band in the lattice with time-reversal symmetry. 
In principle, the half-filled spin-holomorphic FTI studied in this work can be mapped to its lattice version via the basis function mapping between the lowest-Landau-level wave functions and the hybrid Wannier functions~\cite{Qi11,Wu12}.
If so, our study predicts that a strongly correlated, half-filled Chern band with time-reversal symmetry can host a spin-separated FTI, and thus a spin filter.

The spin separation can be directly confirmed by placing the sample in a Kerr microscope at low temperature, where the Kerr rotation measurement~\cite{Kato04} can show the accumulation of opposite spins in two halves of the sample. 
The quantization of the spin Hall conductivity can be verified either in a spin-filtered experiment or in a charge transport experiment by measuring the four-terminal resistance, as shown by K\"{o}nig {\it et al.}~\cite{Konig07}. 
Lastly, the observation of an activated longitudinal resistance can establish the incompressibility of the half-filled spin-holomorphic state.

%%%%%%%%
%%%%%%%%
{\bf Methods}
%%%%%%%%
%%%%%%%%

%%%%%%%%%%%%%%%%%
{\bf Coulomb matrix elements.} 
%%%%%%%%%%%%%%%%%
In the spin-holomorphic Landau levels, the interaction Hamiltonian can be written in second quantization as follows:
\begin{align}
H &= \sum_{m_1,m_2,m_3,m_4} c_{m_1\uparrow}^{\dagger} c_{m_2\uparrow}^{\dagger} c_{m_4\uparrow} c_{m_3\uparrow} \langle m_1, m_2 | V | m_3, m_4 \rangle
\nonumber \\
&+\sum_{m_1,m_2,m_3,m_4} \bar{c}_{m_1\downarrow}^{\dagger} \bar{c}_{m_2\downarrow}^{\dagger} \bar{c}_{m_4\downarrow} \bar{c}_{m_3\downarrow} \langle \overline{m}_1, \overline{m}_2 | V | \overline{m}_3, \overline{m}_4 \rangle
\nonumber \\
&+\sum_{m_1,m_2,m_3,m_4} c_{m_1\uparrow}^{\dagger} \bar{c}_{m_2\downarrow}^{\dagger} \bar{c}_{m_4\downarrow} c_{m_3\uparrow} \langle m_1,\overline{m}_2 | V | m_3,\overline{m}_4 \rangle,
\label{eq:Full_Hamiltonian}
\end{align}
where $c^\dagger_{m\uparrow}$ and $\bar{c}^\dagger_{m\downarrow}$ are the respective creation operators for spin up and down electrons in the holomorphic and antiholomorphic orbitals, respectively, with quantum number $m$, which is the absolute value of the $z$-component angular momentum eigenvalue, $|l_z|$. 
Note that $l_z \geq 0$ and $\leq 0$ for spin up and down electrons in the spin-holomorphic Landau levels, respectively.
The first two terms in Eq~\eqref{eq:Full_Hamiltonian} are exactly the same as those in the usual quantum Hall systems. 
What is different in the spin-holomorphic systems is the last term, which describes the interaction between different spins with spin-dependent holomorphicity.

Concretely, in the spin-holomorphic disc geometry, the Coulomb matrix elements between spin-up electrons are written as follows:
\begin{align}
&\langle m_1,m_2|{ V(|{\bf r}_1-{\bf r}_2|)}| m_3,m_4 \rangle \nonumber \\
&= \int d^2 {\bf k} \tilde{V}_{k} 
\langle m_1,m_2| e^{i {\bf k} \cdot ({\bf r}_1-{\bf r}_2)} | m_3,m_4 \rangle \nonumber \\
&= \int d^2 {\bf k} \tilde{V}_{k} A_{m_1 m_3}({\bf k}) A_{m_2 m_4}(-{\bf k}) ,
\label{eq:Matrix_element_same1}
\end{align}
where $\tilde{V}_{k}$ is the Fourier component of $V(r)$, and $A_{m m^\prime}({\bf k})=\langle m | e^{i {\bf k}\cdot{\bf r}} | m^\prime \rangle = \int d^2 {\bf r} \phi_m^*({\bf r}) e^{i {\bf k}\cdot{\bf r}} \phi_{m^\prime}({\bf r})$. 
After making use of some analytical properties of the lowest-Landau-level eigenstates, $\phi_m({\bf r})$, it can be shown~\cite{Park00} that $A_{m m^\prime}({\bf k})= (i\kappa)^{m-m^\prime} {\cal L}_{m m^\prime}(k) e^{-k^2/2}$, where ${\cal L}_{m m^\prime}(k)=\sqrt{\frac{2^{m^\prime}m^\prime !}{2^m m!}} L^{m-m^\prime}_{m^\prime}(k^2/2)$ with $\kappa=k_x+ik_y=k e^{i\theta}$ and $L^r_n(x)$ being the generalized Laguerre polynomial. 
Then, due to a nice separation of variables between $k$ and $\theta$, Equation~\eqref{eq:Matrix_element_same1} can be rewritten as follows:
\begin{align}
&\langle m_1,m_2|{ V(|{\bf r}_1-{\bf r}_2|)}| m_3,m_4 \rangle \nonumber \\
&= i^{m_1-m_3} (-i)^{m_2-m_4} \nonumber \\
&\times \int k dk \tilde{V}_{k} {\cal L}_{m_1 m_3}(k) {\cal L}_{m_2 m_4}(k) e^{-k^2} k^{m_1+m_2-m_3-m_4} \nonumber \\
&\times \int d \theta e^{i\theta(m_1+m_2-m_3-m_4)} ,
\label{eq:Matrix_element_same2}
\end{align}
where the last factor imposes a selection rule for the $m$ indices indicating the angular momentum conservation, $m_1+m_2=m_3+m_4$. 
The Coulomb matrix elements between spin-down electrons are exactly the same as those above since they are both real and related to each other via complex conjugation.

Meanwhile, the Coulomb matrix elements between different spins are given as follows:
\begin{align}
&\langle m_1,\overline{m}_2|{ V(|{\bf r}_1-{\bf r}_2|)}| m_3, \overline{m}_4 \rangle \nonumber \\
&= \int d^2 {\bf k} \tilde{V}_{k} 
\langle m_1,\overline{m}_2| e^{i {\bf k} \cdot ({\bf r}_1-{\bf r}_2)} | m_3,\overline{m}_4 \rangle \nonumber \\
&= \int d^2 {\bf k} \tilde{V}_{k} A_{m_1 m_3}({\bf k}) A_{m_4 m_2}(-{\bf k}) \nonumber \\
&= \int d^2 {\bf k} \tilde{V}_{k} A_{m_1 m_3}({\bf k}) A^*_{m_2 m_4}({\bf k}) ,
\label{eq:Matrix_element_different1}
\end{align}
where we have used $A_{m m^\prime}({\bf k})=A^*_{m^\prime m}(-{\bf k})$.
Again, due to a nice separation of variables between $k$ and $\theta$, the above equation can be rewritten as follows:
\begin{align}
&\langle m_1,\overline{m}_2|{ V(|{\bf r}_1-{\bf r}_2|)}| m_3,\overline{m}_4 \rangle \nonumber \\
&= i^{m_1-m_3} (-i)^{m_2-m_4} \nonumber \\
&\times \int k dk \tilde{V}_{k} {\cal L}_{m_1 m_3}(k) {\cal L}_{m_2 m_4}(k) e^{-k^2} k^{m_1+m_2-m_3-m_4} \nonumber \\
&\times \int d \theta e^{i\theta(m_1-m_2-m_3+m_4)} ,
\label{eq:Matrix_element_different2}
\end{align}
where it is important to note that the selection rule for the $m$ indices is now changed to $m_1-m_2=m_3-m_4$.
At first sight, this selection rule may seem as if it breaks the angular momentum conservation law, but that is not true since the actual angular momenta for spin-down electrons are $l_z= -m_2$ and $-m_4$, and therefore it is in fact exactly the angular momentum conservation law. 
The comparison between Eqs.~\eqref{eq:Matrix_element_same2} and \eqref{eq:Matrix_element_different2} tells us that the Coulomb matrix elements between the same and different spins would be exactly the same if it were not for the change in the selection rule.

Fundamentally, this change in the selection rule is the physical origin of the spin separation at half filling of the spin-holomorphic Landau levels.
To understand this, it is important to note that the $m$ indices denote the radial locations of the Landau level eigenstates. 
The peculiar selection rule for the Coulomb matrix elements in the spin-holomorphic Landau levels dictates that, after scattering, electrons with different spins should move together to the same radial direction in order to keep the $m$-index difference the same.
This means that electrons with different spins cannot avoid each other effectively unless there is a spontaneous breaking of the spatial symmetry so that they are spatially separated from the outset and thus have no chance to encounter each other.

%[DO NOT ERASE.]
%Explicitly, the Coulomb matrix elements between spin-up eleetrons are given as follows:
%\begin{align}
%&\langle m_1,m_2|{ V(|{\bf r}_1-{\bf r}_2|)}| m_3,m_4 \rangle \nonumber\\
%&=  \delta _{m_1+m_2, m_3+m_4} \frac{1}{2}
%\left( \frac{{m_3}! {m_4}!}{{m_1}! {m_2} !} \right)^{1/2} \nonumber\\
%&\times \sum_{m=0}^{m_3} \binom{m_1}{m_3-m}  \sum_{m^\prime=0}^{m_4} \binom{m_2}{m_4-m^\prime}  \nonumber\\
%&\times \frac{(-1)^{m_1-m_3+m+m^\prime}}{m! m^\prime!} \frac{\Gamma(m+m^\prime+1/2)}{2^{m+m^\prime}} 
%\end{align}
%Meanwhile, the Coulomb matrix elements between different spins are given as follows:
%\begin{align}
%&\langle m_1,\overline{m}_2|{ V(|{\bf r}_1-{\bf r}_2|)}| m_3,\overline{m}_4 \rangle \nonumber\\
%&=  \delta _{m_1-m_2, m_3-m_4} \frac{1}{2}
%\left( \frac{{m_3}! {m_4}!}{{m_1}! {m_2} !} \right)^{1/2} \nonumber\\
%&\times \sum_{m=0}^{m_3} \binom{m_1}{m_3-m}  \sum_{m^\prime=0}^{m_4} \binom{m_2}{m_4-m^\prime}  \nonumber\\
%&\times \frac{(-1)^{m+m^\prime}}{m! m^\prime!} \frac{\Gamma(m_1-m_3+m+m^\prime+1/2)}{2^{m_1-m_3+m+m^\prime}} ,
%\end{align}
%where $\binom{n}{k}$ is the binomial coefficient, and $\Gamma(z)$ is the gamma function.

The same logic applies to the spin-holomorphic spherical geometry. 
Specifically, in the spin-holomorphic spherical geometry, the Coulomb matrix elements between spin-up electrons (and between spin-down electrons) are given as follows:
\begin{align}
&\langle m_1,m_2 |  V(|{\bf r}_1-{\bf r}_2|)  | m_3,m_4 \rangle  \nonumber\\
&=\int d\Omega_1 \int d\Omega_2 Y^*_{QQm_1}({\bf r}_1) Y^*_{QQm_2}({\bf r}_2)    \frac{1}{|{\bf r}_1-{\bf r}_2|}  \nonumber\\  
&\times Y_{QQm_3}({\bf r}_1)Y_{QQm_4}({\bf r}_2) \nonumber\\ 
&=  \frac{1}{\sqrt{Q}}\sum_{l=0}^{2Q}\sum_{m=-l}^{l}  \langle Q,m_1; l,m | Q, m_3 \rangle \langle Q,m_4; l,m | Q,m_2 \rangle \nonumber\\ 
&\times \langle Q,Q; l,0 | Q, Q \rangle^2,
\label{eq:Matrix_element_same_sphere}
\end{align}
where $Y_{Qlm}$ represents the monopole harmonics with the monopole strength $Q$, the angular momentum quantum number $l$, and the $z$-component angular momentum quantum number $m$.
$\langle j_1, m_1; j_2, m_2 | J, M \rangle$ is the Clebsch-Gordan coefficient. 
Above, we have used the expansion of the Coulomb potential on the surface of a sphere in terms of the spherical harmonics~\cite{Jain_Book}:
\begin{align} 
\frac{1}{|{\bf r}_1-{\bf r}_2|} = \frac{4\pi}{R} \sum_{l=0}^{\infty} \sum_{m=-l}^{l} \frac{1}{2l+1} Y^*_{0lm}(\Omega_1) Y_{0lm}(\Omega_2) ,
\end{align}
where $R$ is the radius of a sphere, which is set equal to $\sqrt{Q}$ as usual. 
Similarly, the Coulomb matrix elements between different spins are given as follows:
\begin{align}
&\langle m_1,\overline{m}_2 |  V(|{\bf r}_1-{\bf r}_2|)  | m_3,\overline{m}_4 \rangle  \nonumber\\
&=\int d\Omega_1 \int d\Omega_2 Y^*_{QQm_1}({\bf r}_1) Y_{QQm_2}({\bf r}_2)    \frac{1}{|{\bf r}_1-{\bf r}_2|}  \nonumber\\  
&\times Y_{QQm_3}({\bf r}_1)Y^*_{QQm_4}({\bf r}_2) \nonumber\\ 
&=  \frac{1}{\sqrt{Q}}\sum_{l=0}^{2Q}\sum_{m=-l}^{l}  \langle Q,m_1; l,m | Q, m_3 \rangle \langle Q,m_2; l,m | Q,m_4 \rangle \nonumber\\ 
&\times \langle Q,Q; l,0 | Q, Q \rangle^2,
\label{eq:Matrix_element_different_sphere}
\end{align}
where it is important to note that the only difference between Eqs.~\eqref{eq:Matrix_element_same_sphere} and \eqref{eq:Matrix_element_different_sphere} is that $m_2$ and $m_4$ are swapped, which changes the selection rule for the $m$ indices from $m_1+m_2=m_3+m_4$ to $m_1-m_2=m_3-m_4$, similar to the spin-holomorphic disc geometry.

%%%%%%%%%%%%%%%%%%%%%%%%%%%%%%%%%%%%%%%%
{\bf Angular momentum operators in the spin-holomorphic spherical geometry.} 
%%%%%%%%%%%%%%%%%%%%%%%%%%%%%%%%%%%%%%%%
As mentioned in the main text, the total angular momentum operator is defined as ${\bf L}_{\rm tot}=\sum_i {\bf L}_i$ so that
\begin{align}
{\bf L}^2_{\rm tot} = \sum_{i,j} \left[ \frac{1}{2} (L_{+,i}L_{-,j}+L_{-,i}L_{+,j}) +L_{z,i}L_{z,j} \right] ,
 \end{align}
where $L_{\pm,i}$ are the angular momentum raising and lowering operators of the $i$-th electron, respectively, while $L_{z,i}$ is the $z$-component angular momentum operator of the same electron.
Concretely, 
\begin{align}
L_{\pm,i} &= e^{\pm i\phi_i} \left( \pm \frac{\partial}{\partial \theta_i} + i \cot{\theta_i} \frac{\partial}{\partial \phi_i} + \frac{\hat Q}{\sin{\theta_i}} \right),
\\
L_{z,i} &= -i \frac{\partial}{\partial \phi_i} ,
\end{align}
where $\theta_i$ and $\phi_i$ are the polar and azimuthal angles of the $i$-th electron.

In the usual spherical geometry with a single magnetic monopole, $\hat{Q}$ is just a number representing the monopole strength.
In the spin-holomorphic situation, however, spin up/down electrons experience opposite effective magnetic fields, which are generated by the respective magnetic monopoles with opposite strengths.
A solution to take care of the spin-dependent monopole strengths is to treat $\hat{Q}$ as an operator. 
Specifically, $\hat{Q} Y_{Qlm}(\theta,\phi)= Q Y_{Qlm}(\theta,\phi)$ and $\hat{Q} Y^*_{Qlm}(\theta,\phi)= -Q Y^*_{Qlm}(\theta,\phi)$.

In this generalization, the angular momentum raising and lowering operators are applied to a single particle basis in the spin-holomorphic Landau level as follows:
\begin{align}
L_\pm |Q,Q,m\rangle &= \sqrt{Q(Q+1)-m(m\pm1)} |Q,Q,m\pm1 \rangle, \\
L_\pm |\overline{Q,Q,m}\rangle &= -\sqrt{Q(Q+1)-m(m\mp1)} |\overline{Q,Q,m\mp1} \rangle, \\
L_z |Q,Q,m\rangle &= m |Q,Q,m\rangle, \\
L_z |\overline{Q,Q,m}\rangle &= -m |\overline{Q,Q,m}\rangle ,
\end{align}
where $\langle \theta, \phi |Q,Q,m\rangle = Y_{QQm}(\theta,\phi)$ and $\langle \theta, \phi |\overline{Q,Q,m}\rangle = Y^*_{QQm}(\theta,\phi)$.
Above, the particle index $i$ is dropped for brevity.

\end{document}